\documentclass[epj]{svjour}

\usepackage{graphics}

\newcommand{\Xc}{\Xi_{cc}^+}
\newcommand{\Lc}{\Lambda_c^+}

\begin{document}

\title{Production properties of the Doubly Charmed Baryons at the large Feynman-X}
\author{Sergey Koshkarev}

\date{\today}

\abstract{
This paper focuses on disagreement between theoretical predictions and the SELEX results of the production properties of Doubly Charmed Baryons. The role of the intrinsic charm mechanism in the SELEX kinematic region is researched. The production ratio of the $\Xc$ baryon in the SELEX kinematic region is presented. The recent experimental results are reviewed.
}

\maketitle

\section{Introduction}

In early 2000's the SELEX collaboration published the first observation of 15.9 signal over $6.1 \pm 0.5$ background events of the doubly charmed baryons in the charged decay mode $\Xc \to \Lc K^- \pi^+$ from $\Lc \to p K^- \pi^+$ (1630 events) sample~\cite{SELEX2002}. Three years later the SELEX collaboration reported an observation of 5.62 signal over $1.38 \pm 0.13$ background events of $\Xc \to p D^+ K^-$ decay mode from 1450 $D^+ \to K^- \pi^+ \pi^+$ decays to complement the previously reported decay~\cite{SELEX2005}. The mass and lifetime also have been measured by SELEX (see Table~\ref{tab:table1}). 
\begin{table*}
  \caption{\label{tab:table1}
This table summarizes the SELEX results on measurements of the mass and lifetime of $\Xc$ baryon.
}
\begin{tabular}{|c|c|c|}
\hline
Source&$\Xc$ mass (MeV/$\mbox{c}^2$)&$\Xc$ lifetime (fs) \\
\hline
SELEX~\cite{SELEX2002}&$3519 \pm 1$&$<33$ at 90\% C.L. \\
SELEX~\cite{SELEX2005}&$3518 \pm 3$& not reported \\
\hline
\end{tabular}
\end{table*}
Unfortunately, the SELEX collaboration did not provide production cross-section of the $\Xc$. Still we are able to get the approximated production properties of the $\Xc$ in the order of magnitude accuracy and compared to that of $\Lc$ baryon and $D^+$ meson in kinematic region $x_F  > 0.4$~\cite{Mattson}:
\[
R_{\Lc} = \frac{\sigma(\Xc) \cdot Br(\Xc \to \Lc K^- \pi^+)}{\sigma(\Lc)} \approx 0.01
\]
using known fragmentation ratio $f(c \to \Lc) = 0.071 \pm 0.003~\mbox{(exp.)}\pm 0.018~\mbox{(br.)}$~\cite{BaBarLambda}  and assuming $Br(\Xc \to \Lc K^- \pi^+) \approx Br(\Lc \to p K^- \pi^+) =  (5.0 \pm 1.3)~\%$~\cite{PDG}, one can obtain the ratio of the production cross-section: \\
\begin{eqnarray*}
\frac{\sigma(\Xc)}{\sigma(c\bar{c})} = R_{\Lc} \cdot \frac{f(c \to \Lc)}{Br(\Xc \to \Lc K^- \pi^+)} \approx 1.4 \cdot 10^{-2}.
\end{eqnarray*}
Similar result can be obtained from:
\[
R_{D^+} = \frac{\sigma(\Xc) \cdot Br(\Xc \to p D^+ K^-)}{\sigma(D^+)} \approx 0.004
\]
using fragmentation ratio $f(c \to D^+) = 0.217 \pm 0.014\mbox{(stat.)} \\ ^{+0.013}_{-0.005}\mbox{(syst.)} ^{+0.014}_{-0.016}\mbox{(br.)}$~\cite{ZEUS} and measured ratio $Br(\Xc \to \Lc K^- \pi^+)/Br(\Xc \to p D^+ K^-) = 0.36 \pm 0.21$~\cite{SELEX2005}, one can obtain:
\begin{eqnarray*}
\frac{\sigma(\Xc)}{\sigma(c\bar{c})} = R_{D^+} \cdot \frac{f(c \to D^+)}{Br(\Xc \to p D^+ K^-)} \approx 4.6 \cdot 10^{-2}.
\end{eqnarray*}
So the approximated ratio of the production cross-section is $\frac{\sigma(\Xc)}{\sigma(c\bar{c})} \sim 10^{-2}$.
Comparing this result with theoretically predicted $\sigma(\Xc) / \sigma(c\bar{c}) \sim 10^{-6} - 10^{-5}$~\cite{KiselevUFN,Berezhnoy97}  production ratio for fixed-target experiments with $\pi$ or proton beam and $P_{beam} \sim 600-800$ GeV, we see that measured ratio is at least $10^3$ times larger than theoretical prediction. This is a huge gap between theory and experiment. In paper~\cite{Koshkarev} it has been shown that the kinematic dependencies change the ratio dramatically and obtained a new prediction for the ratio $\sigma(\Xc)/\sigma(c\bar{c}) \sim 10^{-3}-10^{-2}$ which is compared with the SELEX data. The calculation was done in the perturbative approach, however some other researches~\cite{Chang2006,ChangPRD} of the production properties of the doubly charmed baryons and some earlier papers~\cite{Vogt,Brodsky1980,Vogt1995} on the charm production at large Feynman-X point to us importance of the intrinsic charm mechanism. In this paper we research the role of the intrinsic charm mechanism in the production of the doubly charmed baryons in the SELEX kinematic region.

\section{The ratio of production cross-section of $\Xc$ baryon to double-charm at SELEX}

The SELEX experiment is a fixed-target experiment used the Fermilab charged hyperon beam at 600 GeV/c to produce charm particles in a set of thin foil of Cu or in a diamond and operated in the $x_F > 0.3$ kinematic region. The negative beam composition was about 50\% $\Sigma^-$, 50\% $\pi^-$. The positive beam was 90\% protons. 

\subsection{The production cross-section of $\Xc$}

\subsubsection{The Perturbative approach}

The partonic level $\Xc$ production cross-section as a set of parametric functions was given in Ref.~\cite{Berezhnoy97} in following view:
\begin{eqnarray}
\label{eq:aaa}
\hat{\sigma}_{gg} = 213 \cdot \left( 1 - \frac{4 \cdot m_c}{\sqrt{\hat{s}}} \right)^{1.9}  \left( \frac{4 \cdot m_c}{\sqrt{\hat{s}}} \right)^{1.35}~~\mbox{pb},\\
\label{eq:bbb}
\hat{\sigma}_{q\bar{q}} = 206 \cdot \left( 1 - \frac{4 \cdot m_c}{\sqrt{\hat{s}}} \right)^{1.8}  \left( \frac{4 \cdot m_c}{\sqrt{\hat{s}}} \right)^{2.9}~~\mbox{pb}. 
\end{eqnarray}
The numerical coefficients depend on the model parameters, so coefficients above given for $\hat{\sigma} \sim \alpha_s |R(0)|^2 / m_c^2$, where  $\alpha_s = 0.2$, $R(0) = 0.601~\mbox{GeV}^{3/2}$ and $m_c = 1.7$ GeV. These formulae work for the SELEX energies, but cannot be used for LHC energies (see details in Ref.~\cite{Berezhnoy97}).
Combining Eqs. \ref{eq:aaa}, \ref{eq:bbb} and using CTEQ6L~\cite{CTEQ6L} parametrization for parton distribution functions, we may expect $\Xc$ production cross-section in the kinematic region $x_F >0.4$~\cite{Mattson} to be
\[
\sigma_{pQCD}^{x_F >0.4} (\Xc) \approx 2 \div 3~\mbox{pb}.
\]

\subsubsection{The Intrinsic charm approach}

The probability distribution for quark states in the proton can be written as~\cite{Brodsky1980}:
\[
P | p \rangle \approx P | uud \rangle +  P | uudc \bar{c} \rangle +  P | uudc \bar{c} c \bar{c} \rangle + ...
\]
There are two main approaches to have $\Xc$ in the final state in the proton-proton scattering. Charm-gluon scattering $\hat{\sigma}(cg \to \Xc)$ and fragmentation double charm into doubly charmed baryon $\hat{\sigma}(cc \to \Xc)$. As it was shown in Ref.~\cite{Chang2006}  $\sigma(gg \to \Xc):\sigma(gc \to \Xc):\sigma(cc \to \Xc) \approx 1:(25 \div 93):10^{-4}$  so we can see that $\hat{\sigma}(cc \to \Xc)$ gives too small contribution into the final result.

The production cross-section can be written as follows:
\[
\sigma_{IC}(\Xc) = \int dx_1 dx_2   f_g (x_1, \mu) f_c (x_2, \mu) \hat{\sigma}(x_1, x_2),
\]
where $f_{g,c} (x, \mu)$ is gluon~\cite{CTEQ6L} or intrinsic charm~\cite{Pumplin} distribution functions, $x$ is the ratio of the parton momentum to the momentum of the hadron and $\mu$ is the energy scale of the interaction. Explicit view of $\hat{\sigma}(gc \to \Xc)$ can be found in~\cite{ChangPRD}. Doing calculations in the SELEX kinematic region find:
\[
\sigma_{IC}^{x_F >0.4} (\Xc) \approx 10 \times \sigma_{pQCD}^{x_F >0.4} (\Xc).
\] 
Let us remind the reader that in the full kinematic region for the SELEX energies (see Ref.~\cite{Chang2006}) $\sigma_{IC} (\Xc) \approx (25 \div 93) \times \sigma_{pQCD} (\Xc)$.

\subsection{A pair of charmed quarks production}

As it was already shown in~\cite{Koshkarev} in the SELEX kinematic region  the production cross-section of a pair of charmed quarks is suppressed by factor $10^{-4} - 10^{-3}$. This result can be easily obtained with Monte Carlo tools such as described in Refs.~\cite{CalcHep,MadGraph}. 
Upper limit on the intrinsic charm production cross-section at $\sqrt{s} \approx 20-40$ GeV  can be calculated with following approximation~\cite{Vogt,Ingelman}:
\[
\sigma_{IC}(c \bar{c}) \approx 0.1 \times \sigma_{pQCD}(c \bar{c}).
\]
It shows us that this contribution can be neglected in the calculation in the full kinematic region, but can give a significant contribution at the large Feynman-X. Using intrinsic charm distribution function~\cite{Pumplin} we can re-weight $\sigma_{IC}(c \bar{c})$ so that in the SELEX kinematic region:
\[
\sigma_{IC}(c \bar{c}) \approx 0.01 \cdot \sigma_{pQCD}(c \bar{c}) \approx 10^4~\mbox{pb},
\]
that is at least order more than the perturbative approach predicts.

Finally, using calculated $\Xc$ and charm production cross-sections we will obtain a new ratio of the production the doubly charmed baryon and charm at the SELEX experiment:
\[
\frac{\sigma(\Xc)}{\sigma(c \bar{c})}  \sim 10^{-3}.
\]
Keeping in mind that the charm production cross-section is obtained in an optimistic approximation this ratio should be interpreted as the lower limit. So this ratio is compared with the experimentally measured by the SELEX.

\section{Short review of recent results from Belle and LHCb}
\label{subsec:review}

The Belle experiment~\cite{Belle} presented the upper limit on the $\sigma(e^+e^- \to \Xc X)$ is 82-500 fb for the decay mode with the $\Lc$ at $\sqrt{s} = 10.58$ GeV using 980 $\mbox{fb}^{-1}$. The most realistic calculations~\cite{KiselevUFN,Kiselev94} predict $\sigma(\Xc) \simeq 35 \pm 10$ fb what turns out to be at least twice as less as the given limit.

Another recent result from the LHCb experiment~\cite{LHCb} provides the upper limits at 95\% C.L. on the ratio $\sigma(\Xc) \cdot Br(\Xc \to \Lc K^- \pi^+)/\sigma(\Lc)$ to be $1.5 \times 10^{-2}$ and $3.9 \times 10^{-4}$ for lifetimes 100 fs and 400 fs respectively, for an integrated luminosity of 0.65 $\mbox{fb}^{-1}$. It is compared with result from Ref.~\cite{KiselevUFN,Chang2006,ChangPRD,Gunter} $\sim 10^{-4} - 10^{-3}$. However, the LHCb did not reach the lifetime measured by the SELEX experiment yet.

As we can see from above recent results review the other experiments also do not seem to manifest discrepancy between the theory and the experimental data in the properties measured (e.g. small lifetime of the $\Xc$ is important for interpreting of the LHCb data) by the SELEX experiment.

\section{Summary}

In our paper we researched the role of intrinsic charm mechanism in the production properties of doubly charmed baryons in the SELEX kinematic region.  The intrinsic charm mechanism plays leading role in the production of $\Xc$. Comparing theoretical prediction of the doubly charmed baryon production cross-section and production cross-section of a pair of charmed quarks in the SELEX kinematic region we found no significant discrepancy between the theory and the SELEX data. The latest experimental data also is in consistency with theoretical predictions and the properties measured (lifetime of the $\Xc$) by the SELEX experiment.

\begin{acknowledgement}
The authors would like to thank Prof. Stanley Brodsky for pointing out the importance of the intrinsic charm mechanism and Dr. Alexander Rakitin for his friendly support and proofreading the manuscript.
\end{acknowledgement}

\end{document}